\documentclass[twocolumn,aps,amsfonts,amsmath,prd,nofootinbib,preprintnumbers]
{revtex4-1}
\usepackage[T1]{fontenc}
 \usepackage[latin1]{inputenc}
\usepackage{amssymb}
\usepackage{amsmath}
\usepackage{enumerate}
\usepackage{epsfig}
\usepackage[pdftex]{hyperref}
\usepackage{colortbl}

\def\K{K{\"a}hler}
\newcommand{\be}{\begin{eqnarray}}
\newcommand{\ee}{\end{eqnarray}}
\def\ba{\begin{array}}
\def\ea{\end{array}}
\newcommand{\rf}[1]{(\ref{#1})}

\makeatletter

\def\ba{\begin{eqnarray}}
\def\ea{\end{eqnarray}}
\def\beas{\begin{eqnarray*}}
\def\eeas{\end{eqnarray*}}

\def\sla{\raise.15ex\hbox{$/$}\kern-.57em}

\begin{document}

\title{ \Large\bf Chaotic Inflation in Supergravity\\
after Planck and BICEP2}

\vskip 0.5 cm

\author{Renata Kallosh,$^{1}$ Andrei Linde,$^{1}$ and Alexander Westphal$^{1,2}$} 
\vskip 0.2 cm

\affiliation{$^{1}$SITP and Department of Physics, Stanford University, Stanford, CA 94305, USA\\
$^{2}$Deutsches Elektronen-Synchrotron DESY, Theory Group, D-22603 Hamburg, Germany}

\preprint{DESY-04-066}


\begin{abstract}
We discuss the general structure and observational consequences of some of the simplest versions of chaotic inflation in supergravity in relation to the data by Planck 2013 and BICEP2. We show that minimal modifications to the simplest quadratic potential are sufficient to provide  a controllable  tensor mode signal and a suppression of CMB power at large angular scales.
\end{abstract}

\maketitle

\section{Chaotic inflation: The definition} \label{intro}

In this paper we will discuss the simplest versions of the chaotic inflation scenario  in supergravity,  and  compare    their predictions with  the data from Planck 2013 \cite{Ade:2013uln} and BICEP2  \cite{Ade:2014xna}.  But before discussing this issue, we would like to clarify the definition of ``chaotic inflation'', following the original papers and the book on this scenario \cite{Linde:1983gd,Linde:2005ht}. Indeed, some authors incorrectly identify chaotic inflation with the theories with monomial potentials $\phi^{n}$. But there is nothing chaotic about monomial potentials, so what does this strange name refers to?

The name of this broad class of inflationary theories is related to the 
observation that inflation can be realized even in the theories where the inflaton potential does not have any special features such as local minima or maxima with extraordinary small curvature, and even if the universe was not born in the hot Big Bang. To put it to a proper context, one should compare it to other approaches to inflation.

The first version of a theory of inflationary type was proposed by Starobinsky \cite{Starobinsky:1980te}. 
Instead of attempting to solve the homogeneity and isotropy problems, he assumed that  the universe was homogeneous and isotropic from the very beginning, and emphasized that his scenario was ``the extreme opposite of Misner's initial chaos''   \cite{Starobinsky:1980te}. Thus the main goals of this model were different from the goals of inflation.  The goal was to solve the singularity problem by starting the evolution in a non-singular dS state. However, dS state in his scenario was unstable, with a finite decay time \cite{Mukhanov:1981xt}, and therefore it could not exist at $t \to -\infty$. 

Old and new inflation assumed that the universe initially was in a state of thermal equilibrium at an extremely high temperature, 
and then it supercooled and inflated in a state close to the top of the potential $V(\phi)$ \cite{Guth:1980zm,Linde:1981mu,Albrecht:1982wi}. However, old inflation did not quite work, as pointed out by its author \cite{Guth:1982pn}. New inflation resolved most of the problems of old inflation, but it was also ruled out a year later, for many reasons discussed in \cite{Linde:2005ht}. As Hawking said back in 1988, ``the new inflationary model is now dead as a scientific theory, although a lot of people do not seem to have heard about its demise and write papers as if it were viable''  \cite{Hawking1988}.

The chaotic inflation scenario  \cite{Linde:1983gd} was proposed as an alternative to new inflation, after it was realized that the assumption of the hot Big Bang, high temperature phase transitions and supercooling did not help to formulate a successful inflationary scenario. In fact, in most cases these assumptions, which constituted the standard trademark of old and new inflation, made inflation much more difficult to implement. If, instead, one simply considers the universe with different initial conditions in its different parts (or different universes with different values of fields in each of them), one finds that in many of them inflation may occur. It makes these parts exponentially large, thus producing exponentially large islands of order from the primordial chaos. Hence the name: chaotic inflation.

An important feature of this scenario is its versatility and the broad variety of models where it can be implemented. Examples of chaotic inflation models proposed in 1983-1985 included models with monomial and polynomial potentials, and any other models where the slow roll regime was possible. This regime is possible in small field models, but it is especially easy to achieve in large field models, where one could make simple estimates $V'' \sim V/\phi^{2}$ and $V' \sim V/\phi$. Therefore it was argued that in such models the slow-roll conditions can be easily satisfied for $\phi \gg 1$ \cite{Linde:1983gd}. One notable example of such models has the Higgs-like potential $\sim \lambda(\phi^{2}-v^{2})^{2}$ with $v \gg 1$  \cite{Linde:1984cd}. Models of this type later have been  called ``hilltop inflation'' \cite{Boubekeur:2005zm}. Another example was the supergravity-based version of chaotic inflation with the potential $V \sim a(1-e^{-b\phi})$ \cite{Goncharov:1983mw}. Such models became quite popular lately. In 1983-1985, the Starobinsky model  \cite{Starobinsky:1980te} experienced significant modifications. It was reformulated as a theory $R + aR^{2}$, and initial conditions for inflation in this theory were formulated along the lines of the chaotic inflation scenario \cite{Starobinsky:1983zz,Kofman:1985aw}. This resolved the problem with initial conditions of the original version of this model. In the natural inflation scenario, the authors said that ``our model is closest in spirit to chaotic inflation''  \cite{Freese:1990rb}. The hybrid inflation scenario  \cite{Linde:1991km} was introduced as a specific version of the chaotic inflation scenario. 
Step by step, chaotic inflation replaced new inflation in its role of the main inflationary paradigm. Rather than describing some particular subset of inflationary models, it describes the most general approach to inflationary cosmology, which can easily incorporate ideas of quantum cosmology, eternal inflation, inflationary multiverse, and string theory landscape \cite{DeWitt:1967yk,Vilenkin:1982de,Linde:1984ir,Zeldovich:1984vk,Vilenkin:1984wp,Linde:2004nz,Vilenkin:1983xq,Linde:1986fd,Linde:1986fc,Linde:1993xx,Bousso:2000xa,Kachru:2003aw,Douglas:2003um,Susskind:2003kw}.

But this did not happen overnight. Chaotic inflation was so much different from old and new inflation that for a while it was psychologically difficult to accept. 
Even now, 30 years since the demise of new inflation, most of the college books on physics and astrophysics still describe inflation as exponential expansion in the false vacuum state during cosmological phase transitions with supercooling in Grand Unified Theories. That is why a significant part of the first book on inflation \cite{Linde:2005ht} was devoted to the discussion of new inflation versus chaotic inflation. 
 
By now, this discussion is over, but we have a different kind of problem. Every new model belonging to the general class of chaotic inflation is introduced with its own name. That is why some authors 
invented a different classification of models and  
say that chaotic inflation describes only models with monomial potentials, as opposite, e.g., to the hilltop inflation, natural inflation and hybrid inflation. In this paper we will use the original  definition of chaotic inflation following \cite{Linde:1983gd,Linde:2005ht}.

\section{Chaotic inflation in supergravity}

Historically, the first attempts to building inflationary models have been associated with Grand Unified Theories \cite{Guth:1980zm,Linde:1981mu,Albrecht:1982wi}. Chaotic inflation \cite{Linde:1983gd,Linde:2005ht} made this relation unnecessary. Planck results \cite{Ade:2013uln} are consistent with a broad range of theories with the energy scale well below the GUT scale.
But if the results of BICEP2 and their interpretation  in  \cite{Ade:2014xna} are confirmed, it will imply that the energy density 60 e-foldings from the end of inflation was  $\rho \sim (10^{16}~ {\rm GeV})^{4}$. It could suggest, once again, that inflation is somehow related to GUTs. However, the most probable interpretation of the BICEP2 results involves large field inflation \cite{Lyth:1996im}. For example, if one considers the simplest model of chaotic inflation with a quadratic potential, this number appears as a product of the square of the inflaton mass $m \sim 1.5 \times 10^{13}$ ${\rm GeV}$, and the square of the inflaton field, which was $\sim 3 \times 10^{19} $ ${\rm GeV}$. None of these parameters is close to the GUT scale. Moreover, during the last 60 e-foldings of inflation, the energy density of the inflaton field in this model dropped 60 times. Which one of these values of energy density, if any, should we associate with the GUT energy density?

Another popular idea is to associate inflation with supergravity. However, for a very long time it seemed very difficult to do it. The best attempts  were associated with F-term \cite{F} and D-term hybrid inflation \cite{D}, but the simplest versions of these models lead to an excessively large cosmic strings contribution to perturbations of metric, and too small value of tensor modes. It is possible to resolve these problems, but it is not easy.

Before we proceed with the main topic of this paper, inflation in supergravity interacting with chiral matter multiplets, we would like to discuss shortly the new class of supergravity models discovered recently in \cite{Ferrara:2013rsa} which was  called 
``Minimal Supergravity Models of Inflation.''  As different from the standard Einstein frame supergravity models well known to cosmologists, codified by the \K\, potential $K(z, \bar z)$ and a holomorphic superpotential $W(z)$,
depending on chiral superfields, these new models have a  { \it linear  matter multiplet,  describing massive vector or tensor multiplets}. These models do not have a problem of moduli stabilization, due to the fact that these models have only one real scalar field and are characterized by one real  function ${\cal J}(\phi)$ describing the Jordan frame function corresponding to a superconformal model. In these models, one can obtain the inflaton potential of nearly arbitrary shape, and long as it monotonically grows away from the minimum. For example, the superconformal version of these models with the coupling $e^{-{1\over 3} {\cal J} (\phi)} R$, where ${\cal J} = -{1\over 2} \phi^2$, provides a very simple supersymmetric embedding of the $\phi^2$ chaotic inflation \cite{Ferrara:2013rsa}. This class of supergravity cosmological models deserves further investigation as it was discovered only recently.  In this paper we will study more familiar cosmological models in supergravity where there  are {\it chiral matter superfields} and 
where the inflaton field is one of the scalar fields in  chiral multiplets.

The main problem with inflation in supergravity was that the F-term potential in the models with the simplest  \K\, potential  $\Phi\bar\Phi$ contained the exponential factor $e^{\Phi^{2}}$, which typically made the potentials too steep.
The real progress in this direction began with Ref. \cite{Kawasaki:2000yn}, where the simplest model of chaotic inflation with a quadratic potential was introduced. The basic idea is that instead of considering a minimal \K\, potential  $\Phi\bar\Phi$ for the inflaton field, one may consider the  potential $(\Phi+\bar\Phi)^2/2$. This potential has shift symmetry: It does not depend on the field combination $\Phi-\bar\Phi$. Therefore the dangerous term $e^{ K}$ in the F-term potential, which often makes the inflationary potential too steep, is also independent of $\Phi-\bar\Phi$. This makes the potential flat and suitable for chaotic inflation, with the field $\Phi-\bar\Phi$ playing the role of the inflaton. The flatness of the potential is broken only by the superpotential $mS\Phi$, where $S$ is an additional scalar field, which vanishes along the inflationary trajectory. As a result, the potential in the direction $\Phi-\bar\Phi$ becomes quadratic, as in the simplest version of chaotic inflation. Similarly, one can use the \K\, potential $(\Phi-\bar\Phi)^2/2$, with the field $\Phi+\bar\Phi$ playing the role of the inflaton.

This scenario was substantially generalized in \cite{Kallosh:2010ug,Kallosh:2010xz}.  The generalized scenario describes two scalar fields, $S$ and $\Phi$, with the superpotential 
 \be
{W}= Sf(\Phi) \ ,
\label{cond}
\ee
where $f(\Phi)$ is a real holomorphic function such that $\bar f(\bar \Phi) = f(\Phi)$. Any function which can be represented by Taylor series with real coefficients has this property. The \K \, potential can be  chosen to have functional form
\be\label{Kminus}
K= K((\Phi-\bar\Phi)^2,S\bar S).
\ee
In this case, the \K\, potential does not depend on ${\rm Re}\, \Phi$. Under certain conditions on the \K\, potential, inflation occurs along the direction $S = {\rm Im}\, \Phi = 0$. For $\Phi = (\phi+i\chi)/\sqrt 2$,  the field $\phi$ plays the role of the canonically normalized inflaton field with the potential 
\be
V(\phi) = |f(\phi/\sqrt 2)|^{2}.
\ee
 All scalar fields have canonical kinetic terms along the inflationary trajectory $S = {\rm Im}\, \Phi = 0$.

An alternative formulation of this class of models has
the \K\, potential \be\label{Kplus}
K= K((\Phi+\bar\Phi)^2,S\bar S).
\ee
In this class of models, the \K\, potential does not depend on ${\rm Im }\, \Phi$. The role of the inflaton field is played by the canonically normalized field $\chi$ with the potential
\be
V(\chi) = |f(\chi/\sqrt 2)|^{2}.
\ee
One should also make sure that the real part of this field vanishes during inflation. The simplest way to find a class of functions $f(\Phi)$ which lead to the desirable result is to consider any real holomorphic function $f(\Phi) = \sum_{n} c_{n} \Phi^{n}$, and then make the change of variables $\Phi \to -i\Phi$ there. 

Obviously, in the theory with $K= K((\Phi-\bar\Phi)^2,S\bar S)$ it is easier to formulate the required conditions for the function $f$.
However, as long as we do not consider interactions of the field $\Phi$ to vector fields, which are different for scalars and pseudo-scalars, the two approaches give identical results. For example, the theory with $K= -(\Phi-\bar\Phi)^2/2+S\bar S)$ and $f = \Phi + c\Phi^{2}$ leads to the same inflationary scenario as the theory with $K= (\Phi+\bar\Phi)^2/2+S\bar S)$ and $f = -i \Phi - c\Phi^{2}$. Alternatively, one may consider the function  $f = \Phi - i c\Phi^{2}$, obtained from the previous one by multiplication by $i$. 

\section{On Moduli Stabilization in Supergravity and Superstring Theory}

It is widely accepted that moduli stabilization in supersymmetric theories of gravity often presents a challenge for inflationary model building. At present,  the observational predictions for $n_s$ and $r$ from string theory and from supergravity, associated with inflationary  models, require an increasing level of  precision due to gradually improving flow of experimental data. 

It is therefore instructive to revisit the issue of moduli stabilization in general, in view of the fact that in the section above we have described a special  class of supergravity models, which admits stabilization of moduli for all scalars  except the inflaton. This class of models therefore allows an embedding into a supergravity of a rather general set of bosonic inflationary models, where only one scalar, the inflaton,  is light and the rest is heavy during inflation and these heavy fields do not affect the evolution.

In contrast to these models, let us bring an example of the better racetrack inflationary models \cite{better} with two moduli $T_1$ and $T_2$ where the \K\, potential has a shift symmetry 
\be
K= -2 \ln \Big ((T_1+\bar T_1)^{3/2} - (T_2+\bar T_2)^{3/2}\Big)
\ee
and the superpotential has a standard KKLT form
\be
W=W_0 + A e^{-aT_1} + B e^{-bT_2}
\ee
If instead of solving the 4-scalar evolution equations for this model one would consider the slice of it at ${\rm Re} \, T_1$, ${\rm Re} \, T_2$ being constant, as was recently done by many authors, we would obtain a version of natural inflation due to $\cos$-type dependence on axions ${\rm Im} \,  T_1$, ${\rm Im} \,  T_2$ in the models with KKLT potentials for fixed $ {\rm Re}\, T_i$. In this case it would be possible to find models of that type with significant level of gravity waves $r\sim 0.1$.

However, a detailed investigation in \cite{better} shows that to achieve inflation in these models it is necessary to study numerically a simultaneous evolution of all 4 scalars, two dilatons, ${\rm Re}\, T_i$ and 2 axions, ${\rm Im} \,  T_1$, ${\rm Im} \,  T_2$. Consequently, this inflationary string theory model associated with the certain orientifold of string theory predicts $n_s\approx 0.95$ and tiny level of gravity waves, $r< 10^{-5}$.

Thus we see in this example, that a conjecture about moduli stabilization and or about uplifting in each case has to be supported so that predictions of a given model do not change the value of $r$ on many orders of magnitude, depending on whether such a moduli stabilization conjecture is actually correct for a given choice of a model.

The issue of moduli stabilization is especially important for large field inflation in string theory. For example, in accordance with BICEP2, the Hubble constant  at a certain stage of inflation was $H \sim 10^{14}$ GeV.  According to \cite{Kallosh:2004yh,Kallosh:2007wm}, in the simplest KKLT version of the moduli stabilization this implies that the gravitino mass must be greater than $10^{14}$ GeV, and in the models with large volume stabilization the corresponding bound would be 
$m_{3/2} \geq 3 \times 10^{15}$ GeV \cite{Conlon:2008cj}. This would be at odds with the standard assumptions of SUSY phenomenology. The simplest way to avoid this problem is to use a modification of the KKLT procedure with strong moduli stabilization described in \cite{Kallosh:2004yh}, which allows to implement inflation in string theory compatible with the BICEP2 data \cite{Kallosh:2011qk,Dudas:2012wi}, using the general class of inflationary models of  \cite{Kallosh:2010ug,Kallosh:2010xz}. For other approaches to this problem see e.g. \cite{Ibanez:2014zsa}.

\section{Flat Directions and Non-minimal Coupling}

Now that we know how to avoid the problems related to the term $e^{K}$ in the potential and find flat directions which can be used for inflation, we would like to take a step back and interpret these results from a slightly unusual perspective. 

Part of the recent progress in developing new classes of inflationary models in supergravity 
can be traced back to revisiting and further development of the superconformal approach to supergravity, and to reformulation of the standard supergravity models in terms of the Jordan frame. 
Indeed, standard supergravity models with an arbitrary \K\, potential $K(z, \bar z) $ and superpotential $W(z, \bar z) $ can be presented  in the Jordan frame,  where the scalar curvature dependent  term in the action is
\be
{1\over 2}\, \Omega (z, \bar z) \,  R_ J=  {1\over 2}\, e^{-{1\over 3} K (z, \bar z)} \,  R_ J\ .
\label{Relation} \ee
In this form the frame function $\Omega (z, \bar z)R_J $ displaying  the non-minimal couplings of scalars to curvature is related to the  \K\, potential. This was  explained  in \cite{Ferrara:2010in} where the complete action of supergravity in Jordan frame \rf{Relation} was presented.  We present a short  summary of it relevant to cosmology, as well as some new results, in the Appendix of this paper. 

Note, that the term ${1\over 2}\, e^{-{1\over 3} K (z, \bar z)} \,  R_ J$ represents the non-minimal coupling of the scalar fields to gravity in the Jordan frame. The significance of this fact becomes apparent if we remember that the same \K\ potential $ K (z, \bar z)$, which represents the non-minimal coupling to gravity in the Jordan frame, re-appears in the Einstein frame in the coefficient $e^{ K (z, \bar z)}$ in the expression for the F-term potential. It is exactly the term that was responsible for the appearance of the dangerous coefficient $e^{\Phi^{2}}$ preventing chaotic inflation in the theories with the simplest \K\ potential $\Phi\bar\Phi$.

In the previous section we described the way to use flat directions of the \K\ potential for constructing simple versions of chaotic inflation models. Now we have a translation of these rules to another language: {\it Flat directions in the \K\ potential correspond to the fields which are minimally coupled to gravity in the Jordan frame.}

This does not mean that minimal coupling to gravity is necessary for inflation. In particular, supersymmetric version of the Higgs inflation developed in  \cite{Ferrara:2010in} corresponds to the situation where the inflaton field has strongly non-minimal coupling to gravity. There are many other interesting models where this may happen, but most of the recent efforts have been concentrated on the models with the simplest non-minimal coupling of the type $\Omega (z, \bar z) = 1+\xi f(z\bar z)$, which is equivalently described by the logarithmic \K\ potential, see e.g. \cite{Linde:2011nh,Kallosh:2013tua}. In the Appendix of this paper we will study the relation between general \K\ potentials and the non-minimal coupling of the general type $\Omega (z, \bar z)R_J $. In particular, we will relate the case of the polynomial \K\; potentials to non-minimally coupled supergravity models.
In earlier papers, see for example \cite{Kallosh:2014ona},  we have studied this relation in case of the  logarithmic \K\, potentials only.
In discussing practical applications, in this paper we will concentrate on the simplest case of the minimal coupling, but in the end of the paper we will briefly discuss  what may happen in our models from the
 point of view of their observational consequences if we deviate form this rule.

\section{Examples}

The generality of the functional form of the inflationary potential $V(\phi)$ allows one to describe {\it any} combination of the parameters $n_{s}$ and $r$. Indeed, the potential depends only on the function $f(\Phi)$. One can always Taylor expand it, with real coefficients, in a vicinity of the point corresponding to $N \sim 60$ of e-folds, so that the square of this function will fit any desirable function $V(\phi)$ with an arbitrary accuracy. In fact, one can show that there are {\it many} different choices of $f(\Phi)$ which lead to the same values of $n_{s}$ and $r$. Thus, this rather simple class of models may describe {\it any} set of observational data which can be expressed in terms of these two parameters by an appropriate choice of the function $f(\Phi)$ in the superpotential.  

The simplest example of such theory has $f(\Phi) = m\Phi$, which leads, in the context of the theory with the \K\ potential $K= -(\Phi-\bar\Phi)^2/2+S\bar S$, to the simplest parabolic potential  ${m^{2}\over 2} \phi^{2}$ \cite{Kawasaki:2000yn}. It is interesting to analyze various generalizations of this model.

\begin{figure}[t!]
\begin{center}
\hskip -0.76 cm \includegraphics[scale=0.25]{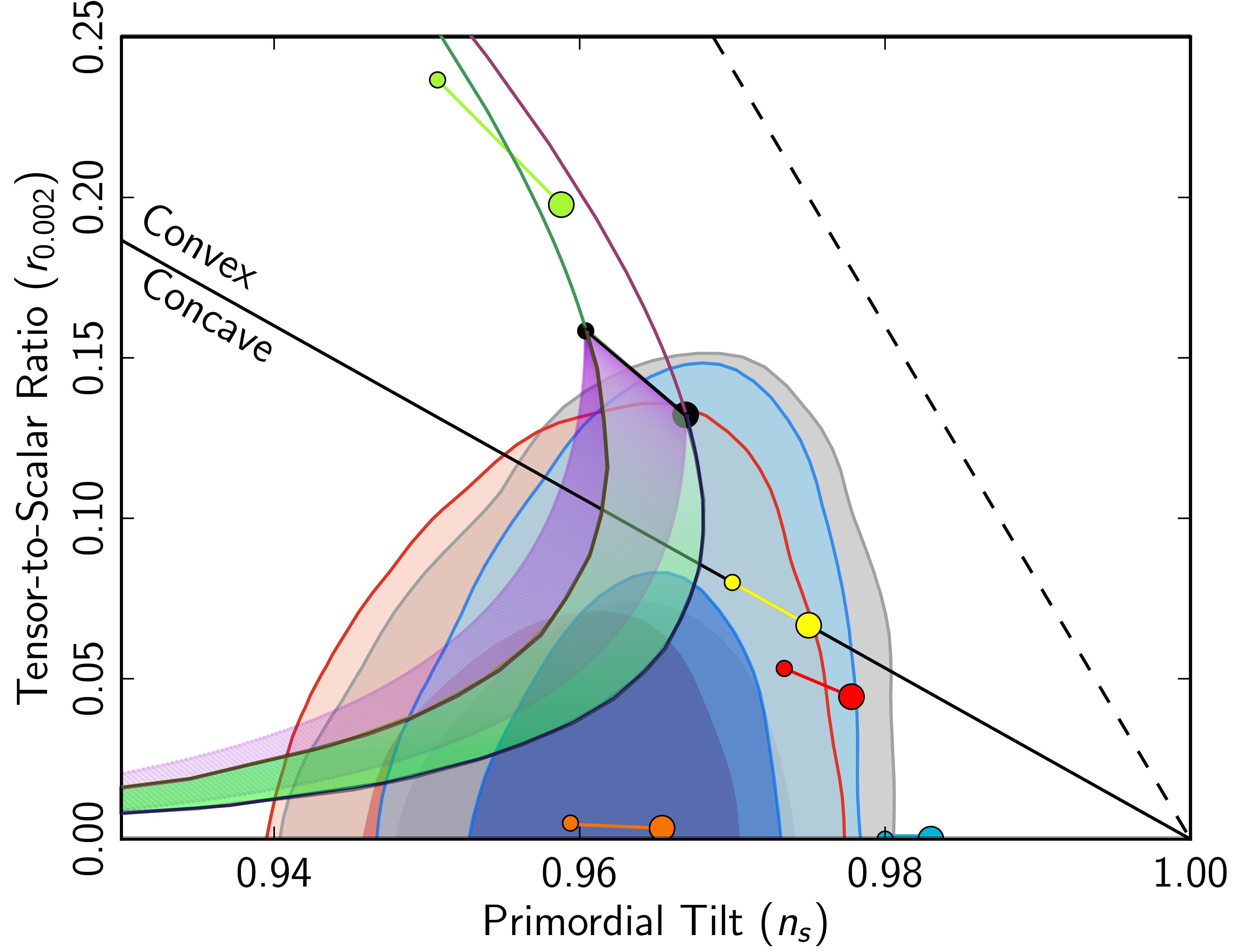}
\end{center}
\caption{\footnotesize The green area describes observational consequences of inflation in the Higgs model \rf{minhiggs} with  $v \gg 1$ ($a \ll 1$), for the inflationary regime when the field rolls down from the maximum of the potential. The continuation of this area upwards corresponds to the prediction of inflation which begins when the field $\phi$ initially is at the slope of the potential at $ |v-\phi| \gg v$. In the limit $v\to \infty$, which corresponds to $a\to 0$, the predictions coincide with the predictions of the simplest chaotic inflation model with a quadratic potential ${m^{2}\over 2} \phi^{2}$.}
\label{chi0}
\end{figure}

As a first step, one may add to the function $f(\Phi) = m\Phi$ a small higher order correction,
\be\label{corr}
f(\Phi) = m\Phi(1-a \Phi)
\ee
with $a \ll 1$. This function, upon the change of variables $\Phi \to \tilde\Phi +{1\over 2a}$, is equivalent to the function $f = -m a (\Phi^{2}-{1\over (2a)^{2}})$ used previously in \cite{Kallosh:2010ug,Linde:2011nh}. Representing $\tilde\Phi$ as $(\phi+i\chi)/\sqrt 2$, one finally obtains the Higgs-type inflationary potential
\be\label{minhiggs}
V(\phi) = {\lambda\over 4}(\phi^{2}-v^{2})^{2}\ , 
\ee
where $\lambda = m^{2} a^{2}$ and $v = {1\over\sqrt 2\, a}$. For $v > 1$, there is an inflationary regime when the field $\phi$ rolls from the maximum of the potential at $\phi = 0$, as in new inflation scenario. Natural initial conditions for inflation in this model are easily set by tunneling from nothing into a universe with spatial topology $T^3$, see e.g.~\cite{Linde:2004nz} and the discussion in~\cite{Linde:2014nna}.  The results of investigation of the observational consequences of this model \cite{Kallosh:2010ug,Kallosh:2007wm,Linde:2011nh} are described by the green area in Figure \ref{chi0}. Predictions of this model are in good agreement with observational data  for a certain range of values of the parameter $a \ll 1$.

However, this does not mean that absolutely any potential $V(\phi)$ can be obtained in this simple context, or that one has a full freedom of choice of the functions $f(\Phi)$. It is important to understand the significance of the restrictions on the form of the \K\ potential and superpotential described above. According to  \cite{Kallosh:2010xz}, in the theory with the \K\ potential $K= K((\Phi-\bar\Phi)^2,S\bar S)$ the symmetry of the \K\ potential $\Phi\to \pm \bar\Phi$, as well as the condition that $f(\Phi)$ is a real holomorphic function are required to ensure that the inflationary trajectory, along which the \K\ potential vanishes is an extremum of the potential in the direction orthogonal to the in\-fla\-ti\-o\-na\-ry trajectory $S = {\rm Im} \Phi = 0$. After that, the proper choice of the \K\, potential can make it not only an extremum, but a minimum,  thus stabilizing the inflationary regime \cite{Kallosh:2010ug,Kallosh:2010xz,Kallosh:2011qk}.

The requirement that $f(\Phi)$ is a real holomorphic function does not affect much the flexibility of choice of the inflaton potential: One can take any positively defined potential $V(\phi)$, take a square root of it, make its Taylor series expansion and thus construct a real holomorphic function which approximate $V(\phi)$ with great accuracy.  However, one should be careful to obey the rules of the game as formulated above.

For example, suppose one wants to obtain a fourth degree polynomial potential of the type of $V(\phi) = {m^{2}\phi^{2}\over 2}(1 + a \phi+ b\phi^{2})$ in supergravity. One may try to do it by taking $K= (\Phi+\bar\Phi)^2/2+S\bar S)$ and $f(\Phi) = m \Phi(1+ c e^{i\theta} \Phi)$ \cite{Nakayama:2013jka}. For general $\theta$, this choice violates our conditions for $f(\Phi)$. In this case, the potential will be a fourth degree polynomial with respect to  $\rm Im\,\Phi$ if $\rm Re\,\Phi = 0$. However, in this model the flat direction of the potential $V(\Phi)$ (and, correspondingly, the inflationary trajectory)   deviate from   $\rm Re\,\Phi = 0$. (Also, in addition to the minimum at $\Phi = 0$, the potential will develop an extra minimum at $\Phi = - c^{{-1}} e^{-i\theta}$.) As a result, the potential along the inflationary trajectory is not exactly polynomial, contrary to the expectations of  \cite{Nakayama:2013jka,Nakayama:2013txa}. Moreover, the kinetic terms of the fields will be non-canonical and non-diagonal.

This may not be a big problem, since the potential in the direction orthogonal to the inflationary trajectory is exponentially steep. Therefore the deviation of this field from $\rm Re\,\Phi = 0$ will not be large, and for sufficiently large values of the inflaton field $\chi$ the potential will be approximately given by the simple polynomial expression $|f(\chi/\sqrt 2)|^{2}$. But in order to make a full investigation of inflation in such models one would need to study evolution of all fields numerically, and make sure that all stability conditions are satisfied. An advantage of the methods developed in \cite{Kallosh:2010ug,Kallosh:2010xz} is that all fields but one vanish during inflation, all kinetic terms are canonical and diagonal along the inflationary trajectory, and investigation of stability is straightforward.

Fortunately, one can obtain an exactly polynomial potential $V(\phi)$ in the theories with $K= K((\Phi-\bar\Phi)^2,S\bar S)$ using the methods of \cite{Kallosh:2010ug,Kallosh:2010xz}, if the polynomial can be represented as a square of a polynomial function $f(\phi)$ with real coefficients. As a simplest example, one may consider $f(\Phi) = m\Phi\bigl(1-c\Phi +d\Phi^{2}\bigr)$. The resulting potential of the inflaton field can be represented as
\be\label{polyn}
V(\phi) = {m^{2}\phi^{2}\over 2}\,\bigl(1-a\phi +a^{2}b\,\phi^{2})\bigr)^{2} \ .
\ee 
Here $a = c/\sqrt 2$ and $a^{2}b = d/2$. We use the parametrization in terms of $a$ and $b$ because it allows us to see what happens with the potential if one changes $a$: If one decreases $a$, the overall shape of the potential does not change, but it becomes stretched. 
The same potential can be also obtained in supergravity with vector or tensor multiplets \cite{Ferrara:2013rsa}. 

\begin{figure}[t!]
\begin{center}
\hskip -0.76 cm \includegraphics[scale=0.42]{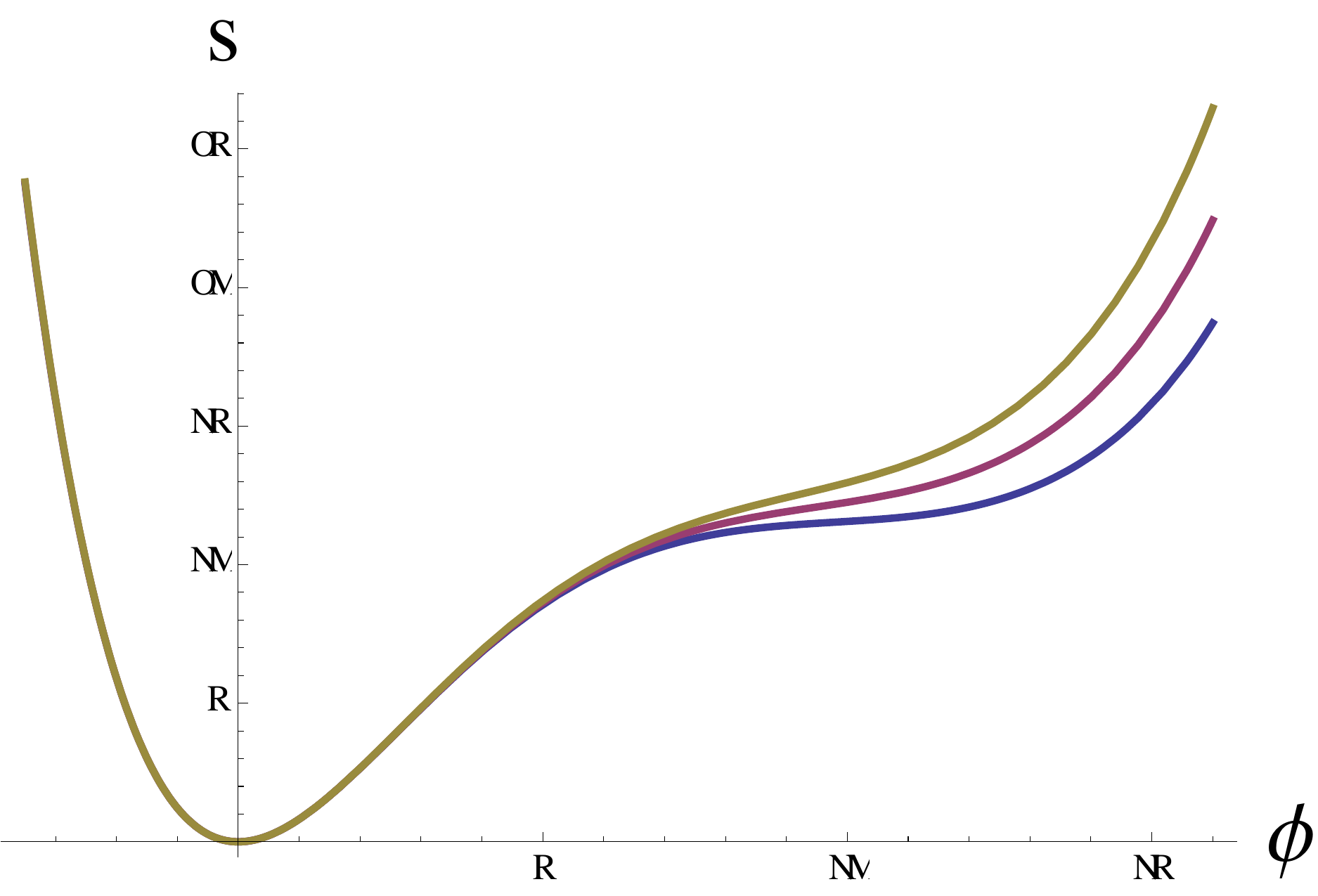}
\end{center}
\caption{\footnotesize The potential $V(\phi) = {m^{2}\phi^{2}\over 2}\,\bigl(1-a\phi +a^{2}b\,\phi^{2})\bigr)^{2}$ for $a = 0.1$ and $b = 0.36$ (upper curve), 0.35 (middle) and 0.34 (lower curve). The potential is shown in units of $m$, with $\phi$ in Planckian units. For each of these potentials, there is a range of values of the parameter $a$ such that the observational predictions of the model are in the region of $n_{s}$ and $r$ preferred by Planck 2013. For $b = 0.34$, the value of the field $\phi$ at the moment corresponding to 60 e-foldings from the end of inflation is $\phi \approx 8.2$. Change of the parameter $a$ stretches the potentials horizontally without changing their shape.}
\label{chi}
\end{figure}

\begin{figure}[t!]
\begin{center}
\hskip -0. cm \includegraphics[scale=0.32]{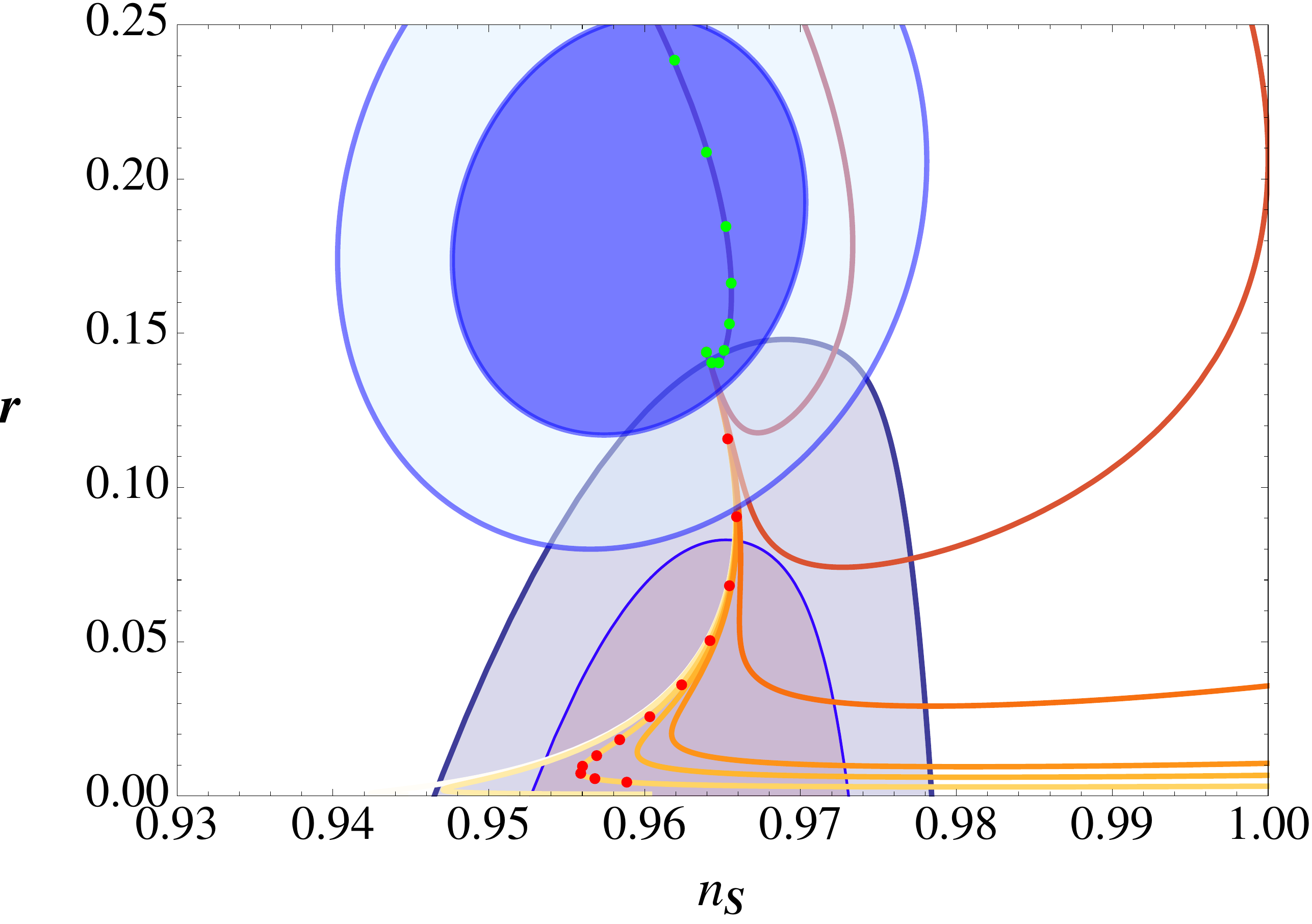}
\end{center}
\vspace*{-3ex}
\caption{\footnotesize Predictions for $n_{s}(a)$ and $r(a)$ in at 55 e-folds the model with $V(\phi) = {m^{2}\phi^{2}\over 2}\,\bigl(1-a\phi +a^{2}b\,\phi^{2})\bigr)^{2}$ for various values of $b = 0.334 \ldots  5$. All curves have $a$ running from $0.001$ to $0.2$. The red ($b=0.34$) and green ($b=5$) balls correspond to $ a = 0.01 \ldots  0.13$ in steps of $0.01$ from the joint start point $a=0.001$ outward. For $a = 0$ one recovers the predictions for the simplest chaotic inflation model with a quadratic potential for all $b$,  while for $b=0.34$ and $a = 0.13$ the predictions almost exactly coincide with the predictions of the Starobinsky model and the Higgs inflation model (red balls). Conversely for $b\gtrsim 1$ and moderately small $a$ our model nicely traverses the BICEP2 constraints within the 1-$\sigma$ area (green balls).}
\label{r-ns-graphs}
\end{figure}

Inflation in this theory may begin under the same initial conditions as in the simplest large field chaotic inflation models $\phi^{n}$. The difference is that in the small $a$ limit, the last 60 e-foldings of inflation are described by the theory $\phi^{2}$. Meanwhile for large $a$ one has the same regime as in the theory $\phi^{6}$, but at some intermediate values of $a$ the last 60 e-foldings of inflation occurs near the point where the potential bends and becomes concave, see Fig. \ref{chi}.  As a result, for $b = 0.34$ and $0.03\lesssim a \lesssim 0.13$ the observational predictions of this model are in perfect agreement with the Planck data, see Fig. \ref{r-ns-graphs}. Agreement with the Planck data can be achieved, for a certain range of $a$, for each of the potentials shown in Fig. \ref{chi}.


We can extend this analysis to a 2-dimensional scan of $a$ and $b$. For a given value of $b$ we saw that plotting $\left(r(a),n_s(a)\right)$ will give a curve in the $(r,n_s)$-plane with a certain segments inside the 1- or 2-$\sigma$ contours, respectively. For this purpose, we approximate the blue-shaded 1- and 2-$\sigma$ contours of the joint PLANCK+WP+BAO exclusion limits on $n_s$ and $r$ from~\cite{Ade:2013uln} (see~Fig.~\ref{chi0}) with a simple polynomial approximation function which is fit to reproduce position, width, height and asymmetric tilt of the PLANCK+WP+BAO exclusion contours. The same way, we approximate the 1- and 2-$\sigma$ contours of the PLANCK+WP+highL+BICEP2 constraints on $n_s$ and $r$~\cite{Ade:2014xna} by slightly asymmetric and rotated ellipses fitting the overall rotation, semi-axes and slight asymmetry of the joint PLANCK+WP+highL+BICEP2 regions. Overlaying these approximated contours with plots of $\left(r(a),n_s(a)\right)$ for a set of values of $b$ in $[0.334\ldots 5]$ gives us Fig.~\ref{r-ns-graphs}. Each curve has $a$ running from $0.001$ to $0.2$.

\begin{figure}[t!]
\begin{center}
\hskip -0. cm \includegraphics[scale=0.42]{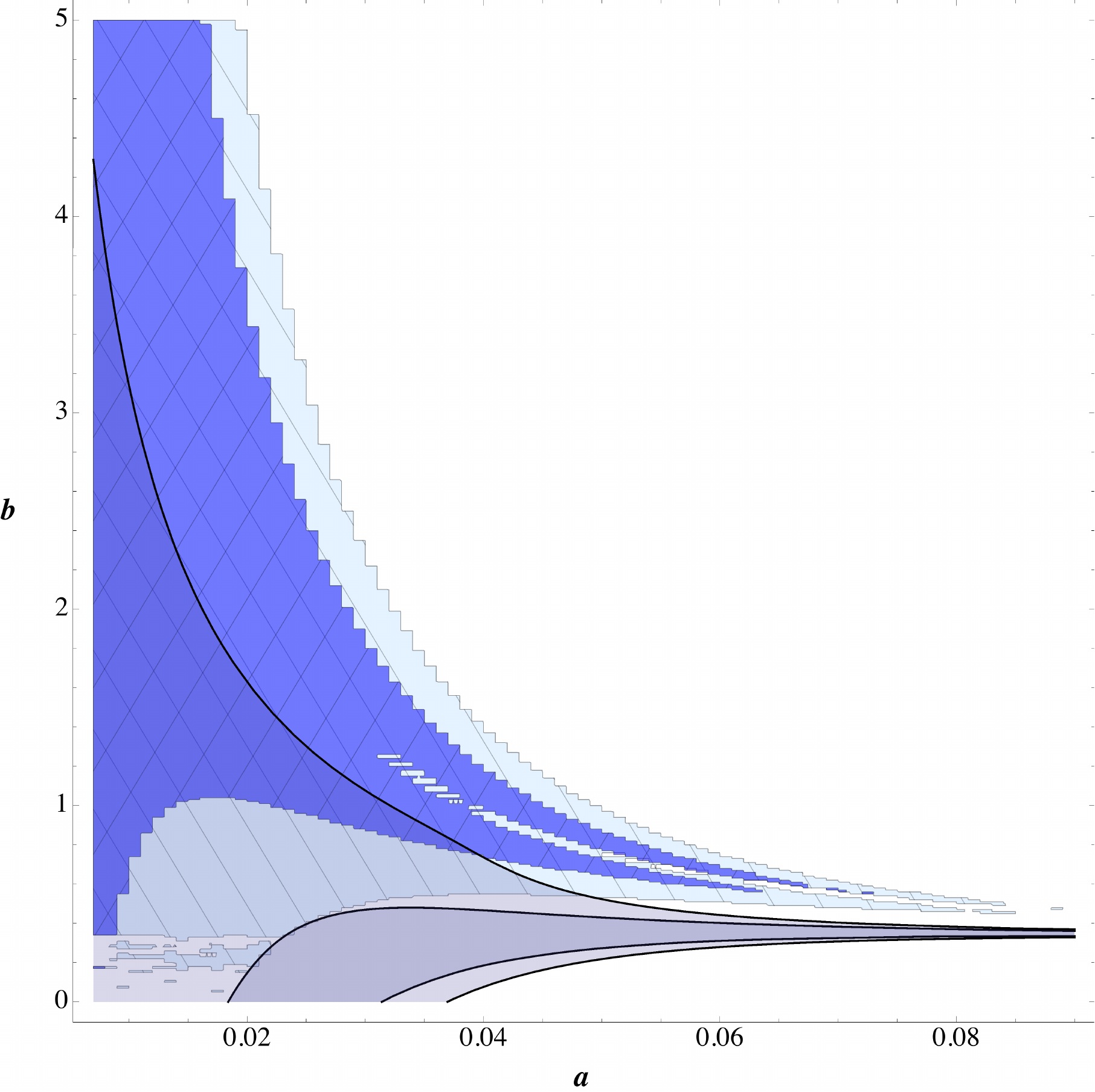}
\end{center}
\caption{\footnotesize {\bf Grey:} Exclusion contours for the microscopic parameters $(a,b)$ in the model with $V(\phi) = {m^{2}\phi^{2}\over 2}\,\bigl(1-a\phi +a^{2}b\,\phi^{2})\bigr)^{2}$ from the PLANCK+WP+BAO approximated exclusion limits on $n_s,r$. Dark grey and light grey denote the $68\%$ and $95\%$ confidence level exclusion contour plot for the microscopic parameters $(a,b)$, respectively, conditioned on $m$ chosen to keep COBE normalization of the curvature perturbation power.\\
{\bf Blue-striped:} Exclusion contours for the microscopic parameters $(a,b)$ in the same model from the BICEP2+PLANCK+WP+highL approximated exclusion limits on $n_s,r$. Dark blue and light blue denote the $68\%$ and $95\%$ confidence level exclusion contour plot for the microscopic parameters $(a,b)$, respectively, conditioned on $m$ chosen to keep COBE normalization of the curvature perturbation power.}
\label{contours-a-b}
\end{figure}

Conversely, using the approximate representation of the PLANCK+WP+BAO exclusion contours, we can numerically solve for the segments of all curves sitting inside the 1- or 2-$\sigma$ contours, respectively. This produces $68\%$ and $95\%$ confidence level exclusion contours for the microscopic parameters $(a,b)$, conditioned on $m$ chosen to keep COBE normalization of the curvature perturbation power. We see the resulting exclusion contours in Fig.~\ref{contours-a-b}. Clearly, compatibility with the data leaves a large fraction of the microscopic parameter space of the model viable, alleviating any perceived need for fine-tuning.

As we see, a slight modification of the simplest chaotic inflation model with a quadratic potential leads to a model consistent with the results of Planck 2013. These results provide us with 3 main data points: 
The amplitude of the perturbations $A_{s}$, the slope of the spectrum $n_{s}$ and the ratio of tensor to scalar perturbations $r$. 
The potential of the model (\ref{polyn}) also depends on 3 parameters which are required to fit the data. Thus we are not talking about fine-tuning where a special combination of many parameters is required to account for a small number of data points; we are trying to fit 3 data points, $A_{s}$,  $n_{s}$ and $r$ by a proper choice of 3 parameters, $m$, $a$ and $b$. The values of $n_{s}$ and $r$ do not depend on the overall scale of $V$; they are fully controlled by the parameters $a$ and $b$. One can show that by fixing a proper combination of $a$ and $b$  with a few percent accuracy, one can cover the main part of the area in the $n_{s} - r$ plane allowed by Planck 2013 and BICEP2. After fixing these two parameters, one can determine the value of $m \sim 10^{{-5}}$ which is required to fit the observed value of $A_{s} \sim 2.2\times 10^{-9}$. 

This is very similar to what happens in the standard model of electroweak interactions, which requires about 20 parameters, which differ from each other substantially. For example, the Higgs coupling to the electron is about $2\times 10^{-6}$. This smallness is required to account for the anomalously small mass of the electron. Meanwhile the Higgs coupling to W and Z bosons and to the top quark are $O(1)$. The cosmological models discussed above are much simpler than the theory of elementary particles. Nevertheless, it would be very nice to identify some possible reasons why the data by WMAP and Planck gradually zoom to some particular area of $n_{s}$ and $r$.


Let us now shortly discuss the effect of the $\Phi^3$ term in the superpotential. For models fitting the CMB data this term is small for field values corresponding to the observable last 50-60 e-folds of inflation. Conversely, the $\Phi^3$ term is the dominant source of shift symmetry breaking in $W$ and thus in the scalar potential at large field values, giving rise to a relatively steep potential $\sim \Phi^6$. Therefore parameter choice $a,b$ giving a good fit to the data will generically consist of an approximately linear or quadratic potential regime up to at least the 60 e-fold value $\phi_{60}$ and then steepen into a sextic potential beyond that point. As mentioned in the conclusions of~\cite{Ade:2013uln}, a similar behavior already observable for a quartic polynomial potential can provide a viable suppression of CMB 2-point function power at large angular scales (low-$\ell$ of $\ell < 40$), for which the PLANCK data provides a hint at about $2.5-3\,\sigma$~\cite{Ade:2013kta}, or including the $r=0.2$ tensor mode contribution~\cite{Ade:2014xna}, at about $3-3.8\,\sigma$~\cite{Bousso:2014jca}.

The same situation arises for our model. Generically, sources of shift symmetry breaking which are subdominant during the observable amount of 
50-60 e-folds of inflation, but lead to rapid steepening of the scalar potential beyond a certain field value, can lead to suppressed CMB power at low-$\ell$. This was discussed in general terms already in~\cite{Linde:1998iw,Linde:2003hc,Contaldi:2003zv}, and more recently in the context of string-inspired models in~\cite{Freivogel:2005vv,Yamauchi:2011qq,Cicoli:2013oba,Pedro:2013pba,Bousso:2013uia,Hazra:2014jka,Freivogel:2014hca,Bousso:2014jca}. A more general discussion of the effects in the large-angular power spectrum of the CMB caused by a generic pre-inflationary phase immediately preceding the observable 50-60 e-folds of inflation will be presented in forthcoming work~\cite{Cicoli:2014xxx}.

\section{Effects of non-minimal coupling to gravity}

\begin{figure}[t!]
\begin{center}
\hskip -0. cm \includegraphics[scale=0.23]{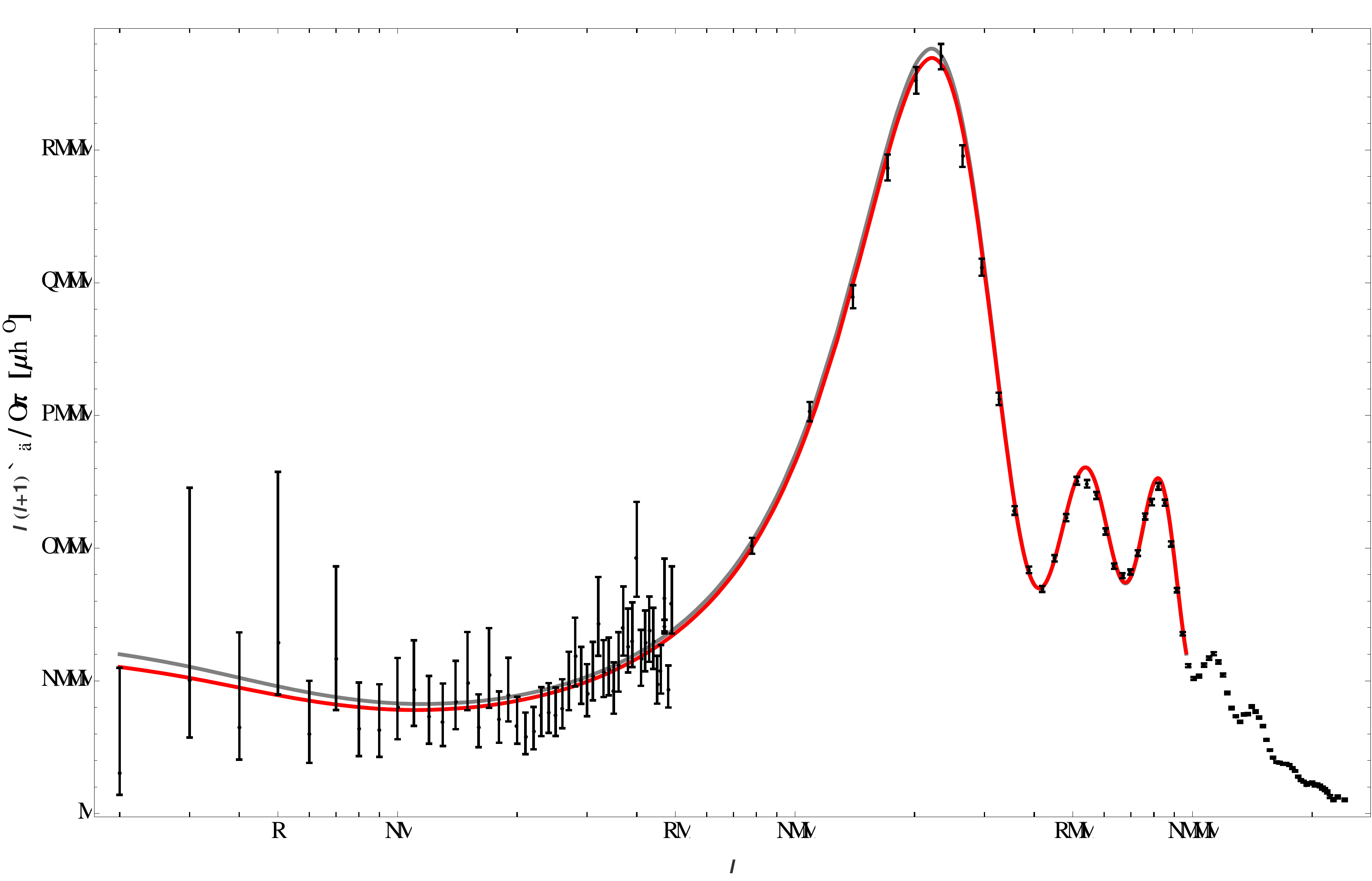}
\end{center}
\caption{\footnotesize Comparison of a model \rf{polyn} with additional steepening from the $\xi$-dependent correction in the K\"abler potential \rf{broken} with the Planck 2013 CMB temperature data. The grey line shows a reference prediction from a $\Lambda$CDM pure power-law power spectrum with $n_s=0.96$. The red line gives the prediction of our model with $a=0.105$ , $b=0.263$ and $\xi=-0.0055$. This model provides $n_s\simeq0.98$ and $r\simeq 0.1$ at $N_e=50$ e-folds before the end of inflation, and generates a power suppression at low-$\ell$ of about 10\%. We see that  {\it the power suppression is confined to a relatively small range of low $\ell$'s}, because the small $\xi$-correction generates an exponentially steep term in the scalar potential at large $\phi$.}
\label{contours-a-b2}
\end{figure}

There are two sources of steepening due to shift symmetry breaking in our class of models. One, as mentioned, is the cubic term in the superpotential, leading to a steep potential $\sim \Phi^{6}$ beyond certain field values. If instead of the superpotential with $f(\Phi) = m\Phi\bigl(1-c\Phi +d\Phi^{2}\bigr)$ one considers a potential with higher powers of $\Phi$, one can achieve a much stronger steepening of the potential. The second possibility of breaking the shift symmetry (as it is already broken in $W$ to get inflation in the first place) arises from giving the inflaton real scalar field a small \emph{negative} non-minimal coupling $\xi$ in the K\"ahler potential. For example, $\xi<0$ modifies the power-law \K\ potential $K= -(\Phi-\bar\Phi)^2/2+S\bar S$ into
\be
 K= -(\Phi-\bar\Phi)^2/2+S\bar S-3\xi\,(\Phi^2+\bar\Phi^2) \ , 
\label{broken}\ee
see eq.~\eqref{Kbr}. For $\xi<0$, this correction to $K$ provides a source of exponentially rapid steeping $e^{3|\xi| \phi^2}$,  which becomes very sharp and pronounced when  $\phi^{2}$ exceeds $1/(3|\xi|)$, see equations \rf{logPot3} and \rf{logPot4}. For the simplest quadratic potential, this effect was studied in \cite{Harigaya:2014qza}, without relating it to the non-minimal coupling. We performed a similar investigation for the polynomial model \rf{polyn}. The results are presented in Fig. \ref{contours-a-b2}. As one can see, the results provide a good fit to observational data, with suppressed  CMB power at low-$\ell$.

\begin{figure}[t!]
\begin{center}
\hskip -0.3 cm \includegraphics[scale=0.37]{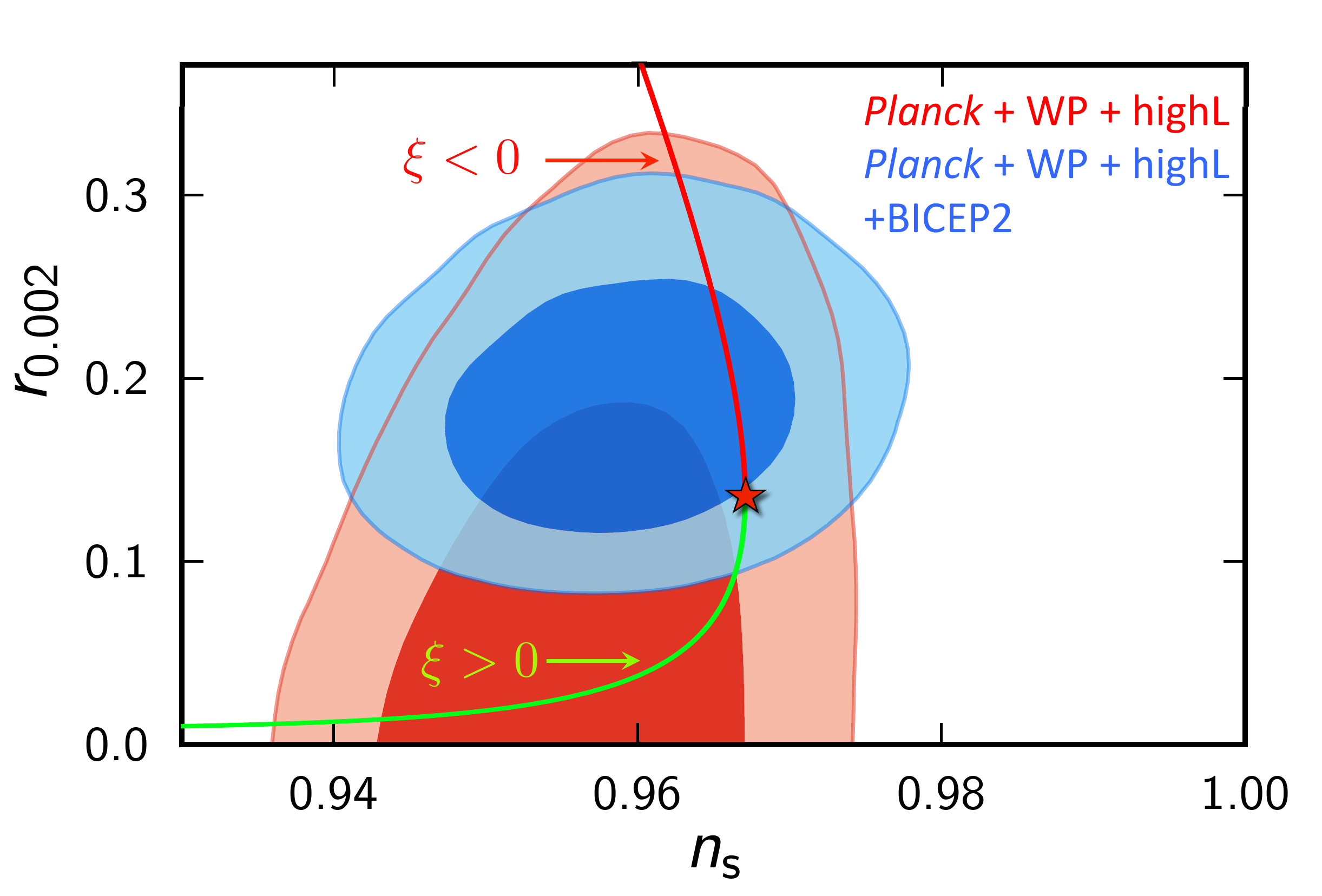}
\end{center}
\vspace*{-3ex}
\caption{\footnotesize Predictions for $n_{s}$ and $r$ for the theory ${m^{2}\over 2} \phi^{2}$ nonminimally coupled to gravity.}
\label{quadrnonmin}
\end{figure}

Finally we should mention another possibility, which was studied in detail in \cite{Linde:2011nh}. If one considers the simplest chaotic inflation model with $V = {m^{2}\over 2}\phi^{2}$ with non-minimal coupling $\Omega = 1 + {\xi} \phi^{2}$, which is described by a supergravity theory with a logarithmic \K\ potential, the predictions of the theory dramatically change even for minuscule deviations of $\xi$ from zero, see Fig. \ref{quadrnonmin}. For $\xi >0$, the value of $r$ becomes sharply lower than at $\xi = 0$, whereas a continuation of $\xi$ to the domain of $\xi <0$ leads to a sharp increase of $r$ while $\xi$ decreases beyond $-O(10^{-3})$ \cite{Linde:2011nh}. 
This demonstrates that non-minimal coupling to gravity is a very powerful tool to control the observational predictions of the theories.  In this context, we should mention a broad class of models with strong non-minimal coupling to gravity \cite{Kallosh:2013tua}, which have the observational predictions continuously interpolating between the predictions of the Starobinsky model with $r \sim 0.004$ and the predictions of the simplest models of chaotic inflation with a broad rangle of values of $r = O(10^{{-1}})$. Further investigation of these possibilities in application to the BICEP2 results is in order.

\section{Conclusions}

In this paper we studied the observational consequences of simplest versions of chaotic inflation in supergravity. As a starting point, we discussed the simplest model with a quadratic potential. 
We can describe this model by the superpotential $W=S\Phi$. Then we perform a very minor modification to this theory by adding to $f(\Phi)$ a tiny term $-a\Phi^{2}$ with $a \ll 1$. The predictions of the model change, as the potential becomes Higgs-type, see equation \rf{minhiggs}. The observational predictions well match Planck data for inflation starting from the top of the potential, and they match BICEP2 data for inflation starting at very large $\phi$. In the limit $a \to 0$, the results coincide with the predictions of the original model with the quadratic potential. Adding yet another term $b\Phi^{3}$ allows us to match Planck/BICEP2 data for various choices of the parameters $a$ and $b$. Finally, adding a small non-minimal coupling to gravity allows to make the potential  sharply rising at large $\phi$, which may suppress CMB power at low-$\ell$.

These are the simplest modifications to the potential that can be consistently implemented in supergravity.  As we already mentioned, the methods developed in \cite{Kallosh:2010ug,Kallosh:2010xz} 
allow us to obtain chaotic inflation in supergravity with inflationary potentials having arbitrary functional shape. 
A recent example, which contains the supergravity realization of natural inflation, is given in \cite{Kallosh:2014vja}. For many years, developing a supergravity version of natural inflation with all non-inflationary moduli stabilized  remained a challenging problem. It was partially resolved in \cite{Kallosh:2007ig,Kallosh:2007cc} under the assumption that one can make the uplifting of the potential in the context of supergravity without perturbing the inflaton potential. The issues of uplifting and moduli stabilization in such models are highly nontrivial. For example, one could expect, naively, that the racetrack inflation \cite{better} should lead to a string theory realization of natural inflation. However, it exhibits an entirely different cosmological dynamics when the evolution of all moduli is taken into account.

Fortunately, using the supergravity theories of the general class developed in \cite{Kallosh:2010ug,Kallosh:2010xz} and  in the present paper, one can stabilize all moduli of the natural inflation scenario and realize it without any need for uplifting \cite{Kallosh:2014vja}. Ref. \cite{Kallosh:2014vja} also describes supergravity versions of chaotic inflation with arbitrary potentials modulated by sinusoidal oscillations, similar to the potentials associated with axion monodromy inflation \cite{Silverstein:2008sg,Kaloper:2008fb,Flauger:2009ab} (see also recent work \cite{Palti:2014kza}).  
Other supergravity models, which can continuously interpolate between the predictions of Planck and BICEP2, can be found in  \cite{Linde:2011nh,Kallosh:2013tua,Kallosh:2013yoa}. These examples show that supergravity offers a  flexible approach to chaotic inflation which allows to match a very broad class of recent observational data.

\section*{Acknowledgments}

We would like to thank V. Mukhanov, K. Nakayama, F. Takahashi, B. Vercnocke and T.T. Yanagida for helpful discussions. RK and AL are supported by the SITP and by the NSF Grant PHY-1316699. AW is supported by the Impuls und Vernetzungsfond of the Helmholtz Association of German Research Centres under grant HZ-NG-603. AW would like to thank SITP,  where the main part of this work was completed, for their warm hospitality.

 \appendix
 
\section{ Inflaton Potentials in Supergravity and 
 Underlying Superconformal Models}

The superconformal models underlying supergravity models,  are defined by the real \K\, potential of the embedding space ${\cal N} (X, \bar X)$ and by the holomorphic superconformal superpotential ${\cal W}(X)$, see for example the textbook \cite{Freedman:2012zz}. The $n+1$ chiral superfields $X$ include all matter superfields as well as a compensator superfield which makes the local superconformal symmetry possible.
In application to cosmology these superconformal models were developed in 
\cite{Kallosh:2000ve} and described recently in a lecture \cite{Kallosh:2014ona}. When the part of the superconformal symmetry which does not belong to supergravity symmetries is spontaneously broken, or gauge-fixed,
the relevant supergravity models with \K\, potential  and holomorphic superpotential,   are derived, both  depend on $n$  chiral superfields $z, \bar z$. In this way a generic Jordan frame supergravity is described. This is useful since  the non-minimal coupling to gravity  plays an interesting role in inflationary models. A superconformal model has the following coupling of scalars to curvature
\be
  e \, {\cal N} (X, \bar X) \, R \ .
\ee
Depending on the choice of the gauge-fixing one derives a Jordan frame supergravity where matter superfields couple to curvature via the so-called frame function $\Omega(z, \bar z)$
\be
 {1\over 2}\,  \Omega (z, \bar z) \,  R_ J\ .
\label{R1} \ee
The corresponding supergravity model in the Einstein frame has the \K\, potential related to the frame function\footnote{The case of a frame function not related to the \K\, potential was also described in \cite{Kallosh:2000ve}, however, the models satisfying \rf{R1} are simplest and therefore often studied. }
\be  
{\cal K} =-3 \log \Omega (z, \bar z) \, , \qquad \Omega (z, \bar z)= \exp ({-{\cal K}/3}) \ .
\label{Kal}\ee 
 There is also a corresponding relation between the superpotential and potential of the superconformal model and that of supergravity. These kind of relations are not unique and reflect that fact that in supergravity there is a \K\, invariance, when the change in the \K\, potential $K(z, \bar z) \rightarrow K(z, \bar z) + g(z) + \bar g(\bar z)$ is compensated by the change of the  superpotential $W(z) \rightarrow W(z) e^{-g(z)}$.

Here we will describe examples of  superconformal models underlying supergravities which are useful for cosmology.

\subsection{General case, polynomial and exponential frame functions}

 Jordan frame supergravity defined by a frame function $\Omega$
 has the following kinetic term for the scalar fields\footnote{There is also an extra kinetic term for scalars from the auxiliary vector fields, which vanishes when scalars are either real or imaginary.}
\be
 \sqrt{-g_J} \,  3 \, \Omega_{\alpha\bar \beta} {\partial}_\mu z^\alpha {\partial}_\nu \bar z^{\bar\beta } g^{\mu\nu}_J \ ,
\ee
where 
\be
\Omega_{\alpha\bar \beta}\equiv {\partial^2 \Omega (z, \bar z)\over \partial z^\alpha \partial \bar z^{\bar \beta}} \ .
\ee 
The inverse matrix $\Omega^{\alpha\bar \beta}$ which one needs to construct the potential  is defined as follows: $\Omega^{\alpha\bar \beta} \Omega_{\bar \beta \gamma}= \delta ^\alpha{}_\gamma$.

In our  cosmological models  \cite{Kallosh:2010ug},
 \cite{Kallosh:2010xz}
 we have two fields $z=(S, \Phi)$. 
The potential in case of 
  ${\cal W}(z, \bar z) = Sf(\Phi)$ in Jordan frame, when $S=0$ is the minimum\footnote{Additional $S$-dependent terms are required for stability at $S=0$, they are described in details in \cite{Kallosh:2010ug},
 \cite{Kallosh:2010xz}}
, is given by the following expression
 \be
V_J= \Omega^{S\bar S} |f(\Phi)|^2 \ .
 \ee
 In this case  the potential in the Einstein frame at $S=0$ is,  
\be
V_E|_{S=0}= {V_J\over \Omega^2}= e^{ {\cal K}(\Phi-\bar \Phi, \Phi+\bar \Phi)}  |f(\Phi)|^2\, .
\label{logPot}\ee
and at $\Phi=\bar \Phi$
\be
V_E|_{S=0, \Phi=\bar \Phi}= {V_J\over \Omega^2}= e^ {{\cal K}(0, 2 \Phi) }  |f(\Phi )|^2\, .
\label{logPot2}\ee
In this case one finds a non-minimal coupling to curvature to the inflaton $ \Phi= \phi/\sqrt 2$ in the form
\be
 {1\over 2}\, e^ {-{1\over 3} {\cal K}(0, \sqrt 2 \phi)  } \,  R_ J\ .
\label{R2} \ee

\subsection{Polynomial frame function, logarithmic \K}
This case was studied in earlier papers  \cite{Kallosh:2010ug},
 \cite{Kallosh:2010xz}
 in detail, namely for the choice 
\be
\Omega_{\rm pol}= 1-{1\over 3}\left(\delta_{\alpha\bar \beta}  z^\alpha \bar z^{\bar \beta}  + J(z) + \bar J(\bar z)\right),
\label{Polframe}\ee
 where $J(z)$ is a holomorphic function of scalars, one has $3 \, \Omega_{\alpha\bar \beta}= - \delta_{\alpha\bar \beta}$, which  means canonical kinetic terms in Jordan frame. This explains why according to eq. \rf{Kal}, we get  a logarithmic \K, when starting with a polynomial frame function. 
\be
{\cal K}_{\rm log} = -3 \log \Big [1-{1\over 3}\left(\delta_{\alpha\bar \beta}  z^\alpha \bar z^{\bar \beta}  + J(z) + \bar J(\bar z)\right)\Big] 
\ee 
For our models with two fields $S, \Phi$  the holomorphic function is
 \be
 J(z)=-{3\chi\over 4 }\Phi^2 \ .
 \ee 
At $\chi=0$ the embedding \K\, manifold defined by the frame function \rf{Polframe} is canonical.  We also define a related non-minimal coupling parameter
 \be
\xi = -{1\over 6} +  {\chi\over 4}.
\label{chi1}\ee 
Thus we find   
\ba
\Omega_{\rm pol} &=& 
1   -{1\over 3}S \bar S + {1\over 6}\left (1+  3\xi \right)  (\Phi-\bar \Phi)^2\nonumber\\ &&+ \,{1\over 2}\xi   (\Phi+\bar \Phi)^2  \ .
\label{AL1} \ea
corresponding to \ba
\mathcal{K}_{\rm log}  &=& -3\log \Bigr[
1   -{1\over 3} S \bar S + {1\over 6}(1+3\xi)  (\Phi-\bar \Phi)^2 \nonumber\\ && +\,{1\over 2}\xi  (\Phi+\bar \Phi)^2 \Bigl].
\label{ALK} \ea

 For cosmological applications the minimum of the potential is at
$S=0$ and $\Phi=\bar \Phi= {\phi\over \sqrt 2}$. Thus during inflation the corresponding superconformal model in the Jordan frame has a negative non-minimal  coupling of the inflaton to the curvature  
 \be
{1\over 2}\,  \Omega_{\rm pol}(\phi)\,  R_J=   {1\over 2}\,  (1-\xi \phi^2) \,  R_J  \ .
\label{R1a} \ee
which explains  the meaning of the parameter $\xi$ as related to non-minimal  coupling of the inflaton to curvature,  $-{\xi\over 2} \phi^2 \,  R$. 

In \cite{Kallosh:2010ug},
 \cite{Kallosh:2010xz} we have studied mostly cosmological models with logarithmic \K\, potentials with $\xi=0$ where the inflaton does not have a non-minimal coupling to the curvature and where $K$ and $W$ and the potential $V_E$ describing the Einstein frame model are
 \ba
\mathcal{K}_{\rm log}  &=& -3\log \Bigr[
1   -{1\over 3} S \bar S + {1\over 6} (\Phi-\bar \Phi)^2  \Bigl].
\label{KP} \ea 
\be
W= Sf(\phi) \ , \qquad V_E|_{S=0, \Phi=\bar \Phi } = |f(\Phi)|^2\ .
\ee
By comparing with \rf{ALK} we note that { \it the shift symmetry of the \K\, potential in \rf{KP} is a consequence of the condition of the minimal coupling of the inflaton to curvature in the underlying superconformal theory}, since at
\be
  \chi=  {2\over 3}\, ,  \qquad \xi=0
\ee
we see in eq. \rf{R1a} that the inflaton is not coupled to curvature.

\

\subsection{Exponential frame function, polynomial \K}
 In \cite{Kallosh:2010ug},
 \cite{Kallosh:2010xz} we have also studied supergravity models which have a polynomial \K\, potential and a shift symmetry.  
 Here  we would like to explain the superconformal origin of a large class of supergravities which have a polynomial \K, rather than logarithmic. We choose the exponential frame function 
 \be
\Omega_{\rm exp}=e^{ -{1\over 3}\left(\delta_{\alpha\bar \beta}  z^\alpha \bar z^{\bar \beta}  + J(z) + \bar J(\bar z)\right)}= \exp ({-{\cal K}/3}) \ ,
\label{frame}\ee
which means that now
\be  
{\cal K}_{\rm pol} =- 3 \log \Omega_{\rm exp} =  \delta_{\alpha\bar \beta}  z^\alpha \bar z^{\bar \beta}  + J(z) + \bar J(\bar z) \, .
\ee  
Here again we focus on  the case with two fields $S, \Phi$ and $J(z)=-{3\chi\over 4 }\Phi^2$. This means that
 \be
\Omega_{\rm exp}=e^ {-{1\over 3}S \bar S + {1\over 6}\left (1+  3\xi \right)  (\Phi-\bar \Phi)^2 +{1\over 2}\xi   (\Phi+\bar \Phi)^2 },
\
\label{frame1}\ee
and 
\begin{equation}
\mathcal{K}_{\rm pol} = (\Phi \bar\Phi + S \bar S)- {3 \chi\over 4 }(\Phi^2+ \bar\Phi^2) .
\label{K1}
\end{equation}
or, using \rf{chi1} we find
\begin{equation}
\mathcal{K}_{\rm pol} = S \bar S - {1\over 2} (\Phi-\bar \Phi)^2 - 3\xi (\Phi^2+ \bar\Phi^2).
\label{Kbr}
\end{equation}
The inflaton potential in the Einstein frame is
\be
V_E|_{S=0, \Phi=\bar \Phi }=  e^ {\cal K_{\rm pol}}  |f(\Phi)|^2= e^{-6\xi \Phi^2}|f(\Phi)|^2 \, .
\label{logPot3}\ee
In terms of  $\Phi=\bar \Phi= {\phi\over \sqrt 2}$ it becomes
\be
V_E|_{S=0, \Phi=\bar \Phi ={\phi\over \sqrt 2}}=  e^{-3\xi \phi^2}|f(\phi/\sqrt 2)|^2 \, .
\label{logPot4}\ee
as shown in eq. (22) of \cite{Kallosh:2010ug}.
In this class of models we find that the non-minimal coupling of the inflaton to curvature at
 $S=0$ and $\Phi=\bar \Phi= {\phi\over \sqrt 2}$ is given by the following expression
 \be
 {1\over 2}\,  \Omega_{\rm exp}  \,  R_J=   {1\over 2} \,e^{ - \xi \phi^2} \,  R_J  \ .
\label{Rexp} \ee
Thus, in this case  in the relevant Jordan frame the inflaton is coupled to curvature exponentially. {\it The deviation of $\xi$ from zero leads to a breaking of the shift symmetry in \K\, potential in} \rf{Kbr}.



\begin{thebibliography}{99}


\bibitem{Ade:2013uln} 
  P.~A.~R.~Ade {\it et al.}  [Planck Collaboration],
``Planck 2013 results. XXII. Constraints on inflation,''
  arXiv:1303.5082 [astro-ph.CO].
  


\bibitem{Ade:2014xna} 
  P.~A.~R.~Ade {\it et al.}  [BICEP2 Collaboration],
  ``BICEP2 I: Detection Of B-mode Polarization at Degree Angular Scales,''
  arXiv:1403.3985 [astro-ph.CO].

   \bibitem{Linde:1983gd} A.~D.~Linde, ``Chaotic Inflation,'' Phys.\ Lett.\ B 
{\bf 129}, 177 (1983). 



\bibitem{Linde:2005ht} 
  A.~D.~Linde,
{\it Particle physics and inflationary cosmology,} (Harwood, Chur, Switzerland, 1990)
  Contemp.\ Concepts Phys.\  {\bf 5}, 1 (1990)
  [hep-th/0503203].



\bibitem{Starobinsky:1980te} 
  A.~A.~Starobinsky,
``A New Type of Isotropic Cosmological Models Without Singularity,''
  Phys.\ Lett.\ B {\bf 91}, 99 (1980).
  
\bibitem{Mukhanov:1981xt} 
  V.~F.~Mukhanov and G.~V.~Chibisov,
``Quantum Fluctuation and Nonsingular Universe. (In Russian),''
  JETP Lett.\  {\bf 33}, 532 (1981)
  [Pisma Zh.\ Eksp.\ Teor.\ Fiz.\  {\bf 33}, 549 (1981)].



\bibitem{Guth:1980zm} 
  A.~H.~Guth,
``The Inflationary Universe: A Possible Solution to the Horizon and Flatness Problems,''
  Phys.\ Rev.\ D {\bf 23}, 347 (1981).

\bibitem{Linde:1981mu} 
  A.~D.~Linde,
``A New Inflationary Universe Scenario: A Possible Solution of the Horizon, Flatness, Homogeneity, Isotropy and Primordial Monopole Problems,''
  Phys.\ Lett.\ B {\bf 108}, 389 (1982).
  
\bibitem{Albrecht:1982wi} 
  A.~Albrecht and P.~J.~Steinhardt,
``Cosmology for Grand Unified Theories with Radiatively Induced Symmetry Breaking,''
  Phys.\ Rev.\ Lett.\  {\bf 48}, 1220 (1982).
  
\bibitem{Guth:1982pn} 
  A.~H.~Guth and E.~J.~Weinberg,
``Could the Universe Have Recovered from a Slow First Order Phase Transition?,''
  Nucl.\ Phys.\ B {\bf 212}, 321 (1983).
   S.~W.~Hawking, I.~G.~Moss and J.~M.~Stewart,
``Bubble Collisions in the Very Early Universe,''
  Phys.\ Rev.\ D {\bf 26}, 2681 (1982).
  

  
  \bibitem{Hawking1988} S.W. Hawking ``A Brief History of Time,'' Bantam Books, NY  (1988).
  
\bibitem{Linde:1984cd} 
 A.~D.~Linde,
``Supergravity And Inflationary Universe. (in Russian),''
  Pisma Zh.\ Eksp.\ Teor.\ Fiz.\  {\bf 37} (1983) 606
   [JETP Lett.\  {\bf 37} (1983) 724].
  A.~D.~Linde,
``Primordial Inflation Without Primordial Monopoles,''
  Phys.\ Lett.\ B {\bf 132}, 317 (1983).
  
\bibitem{Boubekeur:2005zm} 
  L.~Boubekeur and D.~.H.~Lyth,
``Hilltop inflation,''
  JCAP {\bf 0507}, 010 (2005)
  [hep-ph/0502047].

  
\bibitem{Goncharov:1983mw} 
  A.~B.~Goncharov and A.~D.~Linde,
  ``Chaotic Inflation in Supergravity,''
  Phys.\ Lett.\ B {\bf 139}, 27 (1984).
  
\bibitem{Starobinsky:1983zz} 
  A.~A.~Starobinsky,
``The Perturbation Spectrum Evolving from a Nonsingular Initially De-Sitter Cosmology and the Microwave Background Anisotropy,''
  Sov.\ Astron.\ Lett.\  {\bf 9}, 302 (1983).
  
\bibitem{Kofman:1985aw} 
  L.~A.~Kofman, A.~D.~Linde and A.~A.~Starobinsky,
 ``Inflationary Universe Generated by the Combined Action of a Scalar Field and Gravitational Vacuum Polarization,''
  Phys.\ Lett.\ B {\bf 157}, 361 (1985).

\bibitem{Freese:1990rb} 
  K.~Freese, J.~A.~Frieman and A.~V.~Olinto,
``Natural inflation with pseudo - Nambu-Goldstone bosons,''
  Phys.\ Rev.\ Lett.\  {\bf 65}, 3233 (1990).
  
\bibitem{Linde:1991km} 
  A.~D.~Linde,
``Axions in inflationary cosmology,''
  Phys.\ Lett.\ B {\bf 259}, 38 (1991).
  A.~D.~Linde,
``Hybrid inflation,''
  Phys.\ Rev.\ D {\bf 49}, 748 (1994)
  [astro-ph/9307002].
  
\bibitem{DeWitt:1967yk} 
  B.~S.~DeWitt,
``Quantum Theory of Gravity. 1. The Canonical Theory,''
  Phys.\ Rev.\  {\bf 160}, 1113 (1967).

\bibitem{Vilenkin:1982de} 
  A.~Vilenkin,
``Creation of Universes from Nothing,''
  Phys.\ Lett.\ B {\bf 117}, 25 (1982).
  
\bibitem{Linde:1984ir} 
  A.~D.~Linde,
``The Inflationary Universe,''
  Rept.\ Prog.\ Phys.\  {\bf 47}, 925 (1984).
  
\bibitem{Zeldovich:1984vk} 
  Y.~B.~Zeldovich and A.~A.~Starobinsky,
``Quantum creation of a universe in a nontrivial topology,''
  Sov.\ Astron.\ Lett.\  {\bf 10}, 135 (1984).
  
\bibitem{Vilenkin:1984wp} 
  A.~Vilenkin,
``Quantum Creation of Universes,''
  Phys.\ Rev.\ D {\bf 30}, 509 (1984).
  
\bibitem{Linde:2004nz} 
  A.~D.~Linde,
``Creation of a compact topologically nontrivial inflationary universe,''
  JCAP {\bf 0410}, 004 (2004)
  [hep-th/0408164].
  
\bibitem{Vilenkin:1983xq} 
  A.~Vilenkin,
``The Birth of Inflationary Universes,''
  Phys.\ Rev.\ D {\bf 27}, 2848 (1983).
  
\bibitem{Linde:1986fd} 
  A.~D.~Linde,
``Eternally Existing Self-reproducing Chaotic Inflationary Universe,''
  Phys.\ Lett.\ B {\bf 175}, 395 (1986).
  
\bibitem{Linde:1986fc} 
  A.~D.~Linde,
``Eternal Chaotic Inflation,''
  Mod.\ Phys.\ Lett.\ A {\bf 1}, 81 (1986).
  
\bibitem{Linde:1993xx} 
  A.~D.~Linde, D.~A.~Linde and A.~Mezhlumian,
``From the Big Bang theory to the theory of a stationary universe,''
  Phys.\ Rev.\ D {\bf 49}, 1783 (1994)
  [gr-qc/9306035].
  
\bibitem{Bousso:2000xa} 
  R.~Bousso and J.~Polchinski,
``Quantization of four form fluxes and dynamical neutralization of the cosmological constant,''
  JHEP {\bf 0006}, 006 (2000)
  [hep-th/0004134].

  
\bibitem{Kachru:2003aw} 
  S.~Kachru, R.~Kallosh, A.~D.~Linde and S.~P.~Trivedi,
``De Sitter vacua in string theory,''
  Phys.\ Rev.\ D {\bf 68}, 046005 (2003)
  [hep-th/0301240].
  
\bibitem{Douglas:2003um} 
  M.~R.~Douglas,
``The Statistics of string / M theory vacua,''
  JHEP {\bf 0305}, 046 (2003)
  [hep-th/0303194].

  
\bibitem{Susskind:2003kw} 
  L.~Susskind,
``The Anthropic landscape of string theory,''
  In *Carr, Bernard (ed.): Universe or multiverse?* 247-266
  [hep-th/0302219].


  
  

  
\bibitem{Lyth:1996im} 
  D.~H.~Lyth,
``What would we learn by detecting a gravitational wave signal in the cosmic microwave background anisotropy?,''
  Phys.\ Rev.\ Lett.\  {\bf 78}, 1861 (1997)
  [hep-ph/9606387].
  
  \bibitem{F} E.~J.~Copeland, A.~R.~Liddle, D.~H.~Lyth, E.~D.~Stewart and
D.~Wands, ``False vacuum inflation with Einstein gravity,'' Phys.\
Rev.\ D {\bf 49}, 6410 (1994) [astro-ph/9401011]; G.~R.~Dvali,
Q.~Shafi and R.~Schaefer, ``Large scale structure and
supersymmetric inflation without fine tuning,'' Phys.\ Rev.\
Lett.\  {\bf 73}, 1886 (1994) [hep-ph/9406319]

\bibitem{D}
P.~Binetruy and G.~Dvali, ``D-term inflation,'' Phys.\ Lett.\ B
{\bf 388}, 241 (1996) [hep-ph/9606342]; E.~Halyo, ``Hybrid
inflation from supergravity D-terms,'' Phys.\ Lett.\ B {\bf 387},
43 (1996) [hep-ph/9606423].

\bibitem{Ferrara:2013rsa} 
  S.~Ferrara, R.~Kallosh, A.~Linde and M.~Porrati,
``Minimal Supergravity Models of Inflation,''
  Phys.\ Rev.\ D {\bf 88}, 085038 (2013)
  [arXiv:1307.7696 [hep-th]].
  S.~Ferrara, R.~Kallosh, A.~Linde and M.~Porrati,
  ``Higher Order Corrections in Minimal Supergravity Models of Inflation,''
  JCAP {\bf 1311}, 046 (2013)
  [arXiv:1309.1085 [hep-th], arXiv:1309.1085].
  F.~Farakos and R.~von Unge,
 ``Naturalness and Chaotic Inflation in Supergravity from Massive Vector Multiplets,''
  arXiv:1404.3739 [hep-th].







\bibitem{Kawasaki:2000yn}
  M.~Kawasaki, M.~Yamaguchi and T.~Yanagida,
  ``Natural chaotic inflation in supergravity,''
  Phys.\ Rev.\ Lett.\  {\bf 85}, 3572 (2000)
  [arXiv:hep-ph/0004243].
  
  
\bibitem{Kallosh:2010ug} 
  R.~Kallosh and A.~Linde,
  ``New models of chaotic inflation in supergravity,''
  JCAP {\bf 1011}, 011 (2010)
  [arXiv:1008.3375 [hep-th]].
  
\bibitem{Kallosh:2010xz} 
  R.~Kallosh, A.~Linde and T.~Rube,
  ``General inflaton potentials in supergravity,''
  Phys.\ Rev.\ D {\bf 83}, 043507 (2011)
  [arXiv:1011.5945 [hep-th]].

  
  \bibitem{better}
  J.~J.~Blanco-Pillado, C.~P.~Burgess, J.~M.~Cline, C.~Escoda, M.~Gomez-Reino, R.~Kallosh, A.~D.~Linde and F.~Quevedo,
``Inflating in a better racetrack,''
  JHEP {\bf 0609}, 002 (2006)
  [hep-th/0603129].

\bibitem{Kallosh:2004yh} 
  R.~Kallosh and A.~D.~Linde,
``Landscape, the scale of SUSY breaking, and inflation,''
  JHEP {\bf 0412}, 004 (2004)
  [hep-th/0411011];
\bibitem{Kallosh:2007wm} 
  R.~Kallosh and A.~D.~Linde,
  ``Testing String Theory with CMB,''
  JCAP {\bf 0704}, 017 (2007)
  [arXiv:0704.0647 [hep-th]].
  
\bibitem{Conlon:2008cj} 
  J.~P.~Conlon, R.~Kallosh, A.~D.~Linde and F.~Quevedo,
``Volume Modulus Inflation and the Gravitino Mass Problem,''
  JCAP {\bf 0809}, 011 (2008)
  [arXiv:0806.0809 [hep-th]].
  
\bibitem{Kallosh:2011qk} 
  R.~Kallosh, A.~Linde, K.~A.~Olive and T.~Rube,
``Chaotic inflation and supersymmetry breaking,''
  Phys.\ Rev.\ D {\bf 84}, 083519 (2011)
  [arXiv:1106.6025 [hep-th]].
  
\bibitem{Dudas:2012wi} 
  E.~Dudas, A.~Linde, Y.~Mambrini, A.~Mustafayev and K.~A.~Olive,
``Strong moduli stabilization and phenomenology,''
  Eur.\ Phys.\ J.\ C {\bf 73}, 2268 (2013)
  [arXiv:1209.0499 [hep-ph]].
  
\bibitem{Ibanez:2014zsa} 
  L.~E.~Ibanez and I.~Valenzuela,
``BICEP2, the Higgs Mass and the SUSY-breaking Scale,''
  arXiv:1403.6081 [hep-ph].
   
  
\bibitem{Ferrara:2010in}
  S.~Ferrara, R.~Kallosh, A.~Linde, A.~Marrani and A.~Van Proeyen,
``Superconformal Symmetry, NMSSM, and Inflation,''
  arXiv:1008.2942 [hep-th].
  
\bibitem{Linde:2011nh} 
  A.~Linde, M.~Noorbala and A.~Westphal,
``Observational consequences of chaotic inflation with nonminimal coupling to gravity,''
  JCAP {\bf 1103}, 013 (2011)
  [arXiv:1101.2652 [hep-th]].

  
\bibitem{Kallosh:2013tua} 
  R.~Kallosh, A.~Linde and D.~Roest,
``A universal attractor for inflation at strong coupling,''
  Phys.\ Rev.\ Lett.\  {\bf 112}, 011303 (2014)
  [arXiv:1310.3950 [hep-th]].

\bibitem{Kallosh:2014ona} 
  R.~Kallosh,
  ``Planck 2013 and Superconformal Symmetry,''
  arXiv:1402.0527 [hep-th].
  
  
\bibitem{Linde:2014nna} 
  A.~Linde,
``Inflationary Cosmology after Planck 2013,''
  arXiv:1402.0526 [hep-th].




  
  
  
  
  
 

  


  
\bibitem{Nakayama:2013jka} 
  K.~Nakayama, F.~Takahashi and T.~T.~Yanagida,
``Polynomial Chaotic Inflation in the Planck Era,''
  Phys.\ Lett.\ B {\bf 725}, 111 (2013)
  [arXiv:1303.7315 [hep-ph]].
  
 
  
\bibitem{Nakayama:2013txa} 
  K.~Nakayama, F.~Takahashi and T.~T.~Yanagida,
``Polynomial Chaotic Inflation in Supergravity,''
  JCAP {\bf 1308}, 038 (2013)
  [arXiv:1305.5099, arXiv:1305.5099 [hep-ph]].
  

  

  
 

\bibitem{Ade:2013kta} 
  P.~A.~R.~Ade {\it et al.}  [Planck Collaboration],
 ``Planck 2013 results. XV. CMB power spectra and likelihood,''
  arXiv:1303.5075 [astro-ph.CO].
  
\bibitem{Bousso:2014jca} 
  R.~Bousso, D.~Harlow and L.~Senatore,
  ``Inflation After False Vacuum Decay: New Evidence from BICEP2,''
  arXiv:1404.2278 [astro-ph.CO].

  
  
\bibitem{Linde:1998iw} 
  A.~D.~Linde,
``A Toy model for open inflation,''
  Phys.\ Rev.\ D {\bf 59}, 023503 (1999)
  [hep-ph/9807493].

  
\bibitem{Linde:2003hc} 
  A.~D.~Linde,
``Can we have inflation with Omega > 1?,''
  JCAP {\bf 0305}, 002 (2003)
  [astro-ph/0303245].

\bibitem{Contaldi:2003zv} 
  C.~R.~Contaldi, M.~Peloso, L.~Kofman and A.~D.~Linde,
  ``Suppressing the lower multipoles in the CMB anisotropies,''
  JCAP {\bf 0307}, 002 (2003)
  [astro-ph/0303636].
  
\bibitem{Freivogel:2005vv} 
  B.~Freivogel, M.~Kleban, M.~Rodriguez Martinez and L.~Susskind,
 ``Observational consequences of a landscape,''
  JHEP {\bf 0603}, 039 (2006)
  [hep-th/0505232].

\bibitem{Yamauchi:2011qq} 
  D.~Yamauchi, A.~Linde, A.~Naruko, M.~Sasaki and T.~Tanaka,
``Open inflation in the landscape,''
  Phys.\ Rev.\ D {\bf 84}, 043513 (2011)
  [arXiv:1105.2674 [hep-th]].



\bibitem{Cicoli:2013oba} 
  M.~Cicoli, S.~Downes and B.~Dutta,
  ``Power Suppression at Large Scales in String Inflation,''
  JCAP {\bf 1312}, 007 (2013)
  [arXiv:1309.3412 [hep-th], arXiv:1309.3412].
  
\bibitem{Pedro:2013pba} 
  F.~G.~Pedro and A.~Westphal,
  ``Low-$\ell$ CMB power loss in string inflation,''
  JHEP {\bf 1404}, 034 (2014)
  [arXiv:1309.3413 [hep-th]].
  
\bibitem{Bousso:2013uia} 
  R.~Bousso, D.~Harlow and L.~Senatore,
  ``Inflation after False Vacuum Decay: Observational Prospects after Planck,''
  arXiv:1309.4060 [hep-th].
  
\bibitem{Hazra:2014jka} 
  D.~K.~Hazra, A.~Shafieloo, G.~F.~Smoot and A.~A.~Starobinsky,
``Whipped inflation,''
  arXiv:1404.0360 [astro-ph.CO].
  
\bibitem{Freivogel:2014hca} 
  B.~Freivogel, M.~Kleban, M.~R.~Martinez and L.~Susskind,
 ``Observational Consequences of a Landscape: Epilogue,''
  arXiv:1404.2274 [astro-ph.CO].

  


\bibitem{Cicoli:2014xxx} 
  M.~Cicoli, S.~Downes and B.~Dutta, F.~G.~Pedro and A.~Westphal
  ``work in progress''.


  
   
   
    
\bibitem{Harigaya:2014qza} 
  K.~Harigaya and T.~T.~Yanagida,
  ``Discovery of Large Scale Tensor Mode and Chaotic Inflation in Supergravity,''
  arXiv:1403.4729 [hep-ph].
  
  
\bibitem{Kallosh:2014vja} 
  R.~Kallosh, A.~Linde and B.~Vercnocke,
``Natural Inflation in Supergravity and Beyond,''
  arXiv:1404.6244 [hep-th].
  
\bibitem{Kallosh:2007ig} 
  R.~Kallosh,
  ``On inflation in string theory,''
  Lect.\ Notes Phys.\  {\bf 738}, 119 (2008)
  [hep-th/0702059 [HEP-TH]].
  
\bibitem{Kallosh:2007cc} 
  R.~Kallosh, N.~Sivanandam and M.~Soroush,
 ``Axion Inflation and Gravity Waves in String Theory,''
  Phys.\ Rev.\ D {\bf 77}, 043501 (2008)
  [arXiv:0710.3429 [hep-th]].
  
   

 
\bibitem{Silverstein:2008sg} 
  E.~Silverstein and A.~Westphal,
``Monodromy in the CMB: Gravity Waves and String Inflation,''
  Phys.\ Rev.\ D {\bf 78}, 106003 (2008)
  [arXiv:0803.3085 [hep-th]].
  L.~McAllister, E.~Silverstein and A.~Westphal,
``Gravity Waves and Linear Inflation from Axion Monodromy,''
  Phys.\ Rev.\ D {\bf 82}, 046003 (2010)
  [arXiv:0808.0706 [hep-th]].

\bibitem{Kaloper:2008fb} 
  N.~Kaloper and L.~Sorbo,
``A Natural Framework for Chaotic Inflation,''
  Phys.\ Rev.\ Lett.\  {\bf 102}, 121301 (2009)
  [arXiv:0811.1989 [hep-th]].
  N.~Kaloper, A.~Lawrence and L.~Sorbo,
  ``An Ignoble Approach to Large Field Inflation,''
  JCAP {\bf 1103}, 023 (2011)
  [arXiv:1101.0026 [hep-th]].
  
\bibitem{Flauger:2009ab} 
  R.~Flauger, L.~McAllister, E.~Pajer, A.~Westphal and G.~Xu,
  ``Oscillations in the CMB from Axion Monodromy Inflation,''
  JCAP {\bf 1006}, 009 (2010)
  [arXiv:0907.2916 [hep-th]].
  R.~Easther and R.~Flauger,
``Planck Constraints on Monodromy Inflation,''
  JCAP {\bf 1402}, 037 (2014)
  [arXiv:1308.3736 [astro-ph.CO]].
  T.~Kobayashi, O.~Seto and Y.~Yamaguchi,
``Axion monodromy inflation with sinusoidal corrections,''
  arXiv:1404.5518 [hep-ph].


\bibitem{Palti:2014kza} 
  E.~Palti and T.~Weigand,
 ``Towards large r from [p,q]-inflation,''
  arXiv:1403.7507 [hep-th].
  N.~Kaloper and A.~Lawrence,
``Natural Chaotic Inflation and UV Sensitivity,''
  arXiv:1404.2912 [hep-th].
  F.~Marchesano, G.~Shiu and A.~M.~Uranga,
 ``F-term Axion Monodromy Inflation,''
  arXiv:1404.3040 [hep-th].
  A.~Hebecker, S.~C.~Kraus and L.~T.~Witkowski,
``D7-Brane Chaotic Inflation,''
  arXiv:1404.3711 [hep-th].
  T.~Higaki and F.~Takahashi,
  ``Natural and Multi-Natural Inflation in Axion Landscape,''
  arXiv:1404.6923 [hep-th].
  S.~-H.~H.~Tye and S.~S.~C.~Wong,
  ``Helical Inflation and Cosmic Strings,''
  arXiv:1404.6988 [astro-ph.CO].
  R.~Kappl, S.~Krippendorf and H.~P.~Nilles,
  ``Aligned Natural Inflation: Monodromies of two Axions,''
  arXiv:1404.7127 [hep-th].
  I.~Ben-Dayan, F.~G.~Pedro and A.~Westphal,
  ``Hierarchical Axion Inflation,''
  arXiv:1404.7773 [hep-th].
  C.~Long, L.~McAllister and P.~McGuirk,
  ``Aligned Natural Inflation in String Theory,''
  arXiv:1404.7852 [hep-th].
  K.~-Y.~Choi and B.~Kyae,
  ``Large tensor spectrum of BICEP2 in the natural SUSY hybrid model,''
  arXiv:1404.7855 [hep-th].
  
\bibitem{Kallosh:2013yoa} 
  R.~Kallosh, A.~Linde and D.~Roest,
 ``Superconformal Inflationary $\alpha$-Attractors,''
  JHEP {\bf 1311}, 198 (2013)
  [arXiv:1311.0472 [hep-th]];
  R.~Kallosh, A.~Linde and D.~Roest, work in progress.
  

  
\bibitem{Freedman:2012zz} 
  D.~Z.~Freedman and A.~Van Proeyen,
``Supergravity,''
  Cambridge, UK: Cambridge Univ. Pr. (2012) 607 p

\bibitem{Kallosh:2000ve} 
  R.~Kallosh, L.~Kofman, A.~D.~Linde and A.~Van Proeyen,
  ``Superconformal symmetry, supergravity and cosmology,''
  Class.\ Quant.\ Grav.\  {\bf 17}, 4269 (2000)
  [Erratum-ibid.\  {\bf 21}, 5017 (2004)]
  [hep-th/0006179].




\end{thebibliography}
\end{document}